\documentclass[letter]{aa} 
\usepackage{graphicx}
\usepackage{natbib}
\usepackage{amssymb}
\usepackage{txfonts}
\begin{document}

\hyphenation{a-na-ly-sis mo-le-cu-lar pre-vious e-vi-den-ce dif-fe-rent pa-ra-me-ters ex-ten-ding a-vai-la-ble ca-li-bra-tion va-ri-a-bi-li-ty con-ti-nu-um}

\title{Tracing jet emission at the base of a high-mass YSO.}
\subtitle{First AMBER/VLTI observations of the Br$\gamma$ emission in IRAS 13481-6124.
\thanks{Based on observations collected at the VLT (ESO Paranal, Chile) with programmes 090.C-0371(B)}}
\author{A. Caratti o Garatti \inst{1}, B. Stecklum \inst{2}, G. Weigelt \inst{3}, D. Schertl \inst{3}, K.-H. Hofmann \inst{3}, S. Kraus \inst{4},
R.D. Oudmaijer \inst{5}, W.J. de Wit \inst{6}, A. Sanna \inst{3}, R. Garcia Lopez \inst{1}, A. Kreplin \inst{4}, T.P. Ray \inst{1}
}

\offprints{A. Caratti o Garatti, \email{alessio@cp.dias.ie}}

\institute{
Dublin Institute for Advanced Studies, School of Cosmic Physics, Astronomy \& Astrophysics Section,
31 Fitzwilliam Place, Dublin 2, Ireland\\ 
\email{alessio@cp.dias.ie}
\and
Th\"uringer Landessternwarte Tautenburg, Sternwarte 5, D-07778 Tautenburg, Germany
\and
Max-Planck-Institut f\"{u}r Radioastronomie, Auf dem H\"{u}gel 69, D-53121 Bonn, Germany
\and
University of Exeter, School of Physics, Stocker Road, Exeter, EX4 4QL, UK
\and
School of Physics and Astronomy, University of Leeds, Leeds, LS2 9JT, UK
\and
ESO-European Organisation for Astronomical Research in the Southern Hemisphere, Alonso de Cordova 3107, Vitacura, Santiago de Chile, Chile
}

\date{Received date; Accepted date}

 
  \abstract
   {
   }
   {To probe the circumstellar environment of IRAS\,13481-6124, a 20\,M$_\sun$ high-mass young stellar object (HMYSO) with a collimated parsec-scale jet and an accretion disc, we 
   investigate the origin of its Br$\gamma$ emission line through NIR interferometry.}
   {We present the first AMBER/VLTI observations of the Br$\gamma$ emitting region in an HMYSO at medium spectral resolution
($ \mathrm R $=1500). 
   }
   {Our AMBER/VLTI observations reveal a spatially and spectrally resolved Br$\gamma$ line in emission with a strong P\,Cygni profile, indicating  
   outflowing matter with a terminal velocity of $\sim$500\,km\,s$^{-1}$. 
   Visibilities, differential phases, and closure phases are detected in our observations within the spectral line and in the adjacent continuum. 
   Both total visibilities (continuum plus line emitting region) and pure-line visibilities indicate that the Br$\gamma$-emitting region is more compact (2--4\,mas in diameter or $\sim$6--13\,au at 3.2\,kpc) 
   than the continuum-emitting region ($\sim$5.4\,mas or $\sim$17\,au). The absorption feature is also spatially resolved at the longest baselines (81 and 85\,m) 
   and has a visibility that is slightly smaller than the continuum-emitting region. 
   The differential phases at the four longest baselines display an `S'-shaped structure across the line, peaking in the blue- and red-shifted high-velocity
   components. The calibrated photocentre shifts are aligned with the known jet axis, i.e they are probably tracing an ionised jet. 
   The high-velocity components (v$_r$$\sim$100--500\,km\,s$^{-1}$) are located far from the source, whereas
   the low-velocity components (0--100\,km\,s$^{-1}$) are observed to be closer, indicating a strong acceleration of the gas flow in the inner 10\,au.  
   Finally, a non-zero closure phase along the continuum is detected. By comparing our observations with the synthetic images of the continuum around 2.16\,$\mu$m, we 
   confirm that this feature originates from the asymmetric brightness distribution of the continuum owing to the inclination of the inner disc.}
   {}
\keywords{stars: formation -- stars:circumstellar matter -- stars: protostars -- stars: massive -- techniques: interferometric -- techniques: high angular resolution}
\titlerunning{AMBER/VLTI MR observations of the Br$\gamma$ in IRAS\,13481-6124}
\authorrunning{A. Caratti o Garatti et al.}
     
\maketitle
%

\section{Introduction}
\label{introduction:sec}

The evolution of young stellar objects (YSOs), over a wide range of masses, seems to be characterised by significant
accretion from a magnetised circumstellar disc and
ejection of collimated jets. Both processes are tightly related: jets can remove excess angular momentum, 
so that some of the disc material can accrete onto the star~\citep[see, e.g.][]{konigl00,pudritz_PPV}. 
Their interplay is relatively well understood and tested in low-mass YSOs, whereas
not much is known about high-mass YSOs (HMYSOs; $M > 8$\,M$_\odot$; O and early B spectral types), as their formation mechanism and evolution are still a matter of debate~\citep[see, e.g.][]{tan}. 
Substantial progress in understanding the formation mechanism of HMYSOs is gained by connecting 
the gas dynamics close to the forming star~\citep[e.g.][]{kraus10,ilee,cesaroni,sanna15} with the parsec-scale jet and outflow emission being driven by the central accreting 
source~\citep[e.g.][]{sanna14,caratti15}.
Moreover, compelling observational evidence for the accretion disc scenario has recently come from the detection of compact discs in Keplerian
rotation~\citep[e.g.][]{kraus10,ilee,cesaroni} and parsec-scale collimated jets driven by HMYSOs~\citep[e.g.][]{stecklum,varricatt,caratti15}. 

\object{IRAS\,13481-6124} ($\alpha$(J2000)=13:51:37.856, $\delta$(J2000)=-61:39:07.52) is among the few HMYSOs where an accretion disc~\citep[][]{kraus10,ilee,boley}, 
an outflow, and a collimated parsec-scale jet~\citep[][]{kraus10,stecklum12,caratti15} have been detected. 
Located at a distance of $\sim$3.2\,kpc, IRAS\,13481-6124 has a bolometric luminosity of 5.7$\times$10$^4$\,L$_\odot$~\citep[][]{lumsden13}. 
By modelling the spectral energy distribution (SED), \citet{grave} inferred an age of $\sim$10$^4$\,yr and a protostellar mass of $\sim$20\,M$_\odot$ (i.e. an O9 ZAMS spectral type). 
NIR interferometric observations with the Very Large Telescope Interferometer (VLTI) in the $K$-band continuum revealed a compact dusty disc~\citep[$\sim$5.4\,mas in diameter, or $\sim$17.3\,au at 3.2\,kpc;][]{kraus10},
tilted by $\sim$45$\degr$ with respect to the plane of the sky. Modelling of the CO band-head emission lines at 2.3\,$\mu$m suggests a disc in Keplerian rotation~\citep[][]{ilee}. 
Perpendicular to the disc, a well-collimated (precession angle$\sim$8$\degr$) parsec-scale jet has also been detected~\citep[with position angle -P.A.- of $\sim$206$\degr$/26$\degr$ east of north, blue- and red-shifted lobes, respectively;][]{caratti15},
being traced by shocked H$_2$ and [\ion{Fe}{ii}] lines. The shocked H$_2$ emission is observed down to $\sim$20\,000\,au from the source~\citep[][]{stecklum12}.
Closer to the source (a few arcseconds), the H$_2$ emission is mostly excited by fluorescence and it is not detected at the smallest spatial scales observed with SINFONI~\citep[$\sim$0.1$\arcsec$ or $\sim$320\,au;][]{stecklum12}. 
NIR spectroscopy also shows bright \ion{H}{i} emission, confined to the HMYSO circumstellar environment~\citep[$\leq$10\,000\,au;][]{stecklum12}.
The Br$\gamma$ line on source shows a shallow 
P\,Cygni profile and high-velocity wings~\citep[several hundred km\,s$^{-1}$;][]{stecklum12}.
Their spectro-astrometric analysis indicates a photocentre shift between line and continuum.
The Br$\gamma$ line profile might then be the product of different kinematical components, which originate from different locations: 
magnetospheric accretion flows, a hot disc atmosphere, strong winds from the stellar surface, or from the inner regions of the disc (within a few au), or an extended
and collimated ionised jet.

To clarify the nature of the different kinematic components of the Br$\gamma$ line and probe the circumstellar environment of IRAS\,13481-6124, we therefore
carried out, for the first time, VLTI/AMBER spectro-interferometric observations of this HMYSO at medium spectral resolution.
This is the first NIR interferometric study that spatially and spectrally resolves an HMYSO in both continuum and Br$\gamma$ line.

Section~\ref{observations:sec} reports our interferometric observations and data reduction, while the interferometric
results are presented in Sect.~\ref{results:sec}. Finally, in Sect.~\ref{discussion:sec}, we discuss the origin of the Br$\gamma$ in IRAS\,13481-6124.

\section{Observations and data reduction}
\label{observations:sec}

IRAS\,13481-6124 was observed on the 28 February 2013 in two runs
at the ESO/VLTI with AMBER~\citep[][]{petrov07} and the UT1-UT2-UT3 telescope configuration. 
The projected baseline lengths extend from $\sim$38\,m to $\sim$85\,m, and their position angle ranges between $\sim$38$\degr$ and
$\sim$62$\degr$, i.e. relatively close to being parallel to the jet P.A.. Details of the observational settings are reported in Table~\ref{tab:obs}.

We used AMBER medium spectral resolution mode in the $K$-band (MR-2.1 mode with nominal $ \mathrm R $ = 1500) 
covering the spectral range from 1.926 to 2.275\,$\mu$m around the Br$\gamma$-line emission (at 2.166\,$\mu$m).
Although IRAS\,13481-6124 in the $K$ band is bright enough ($K$=4.9\,mag) to be observed interferometrically with AMBER and the UTs, its
$H$-band magnitude ($H$=7.6\,mag) exceeds the limit for employing the fringe tracker FINITO~\citep[][]{gai}, which works in that band.
The adopted detector integration time (DIT) was 0.3\,s per interferogram and we integrated on source for about 25 and 30 minutes in the first and second run, respectively.
Owing to the better average seeing of the second run ($\sim$0.75$\arcsec$ vs. $\sim$0.95$\arcsec$) and the slightly longer exposure on source, the second-run data have a higher
S/N ratio. Star HD 103125 was observed before and after the science observations with the same observational settings,
and used as an interferometric calibrator to derive the transfer function.
To reduce our interferograms, we applied our own data reduction software, that is based on the P2VM algorithm~\citep[][]{tatulli07} and that
provides us with wavelength-dependent visibilities, wavelength-dependent differential phases, closure phases, and wavelength-calibrated spectra.
The wavelength calibration was refined using the numerous telluric lines present in the observed wavelength 
range~\citep[for more details on the wavelength calibration method, see][]{weigelt,rebeca15}. We estimate 
an uncertainty in the wavelength calibration of $\sim$1.5\,\AA~($\sim$20\,km\,s$^{-1}$). The spectral resolution measured on the spectrally 
unresolved telluric features around the Br$\gamma$ line is $ \mathrm R \sim$2200 or $\Delta$v $\sim$ 140\,km\,s$^{-1}$.
To convert the observed wavelengths into radial velocities, we used a local standard of rest (LSR) velocity of -37.9\,km\,s$^{-1}$~\citep[][]{lumsden13}. Therefore all the velocities provided
in this paper are with respect to the LSR.

\begin{table*}
 \centering
\caption{\label{tab:obs} Log of the VLTI/AMBER observations of IRAS\,13481-6124.}
\begin{scriptsize}
\begin{tabular}{ccccccccccc}
\hline\hline
IRAS 13481-6124   & \multicolumn{2}{c}{Time [UT]}& Unit Telescope & Spectral & Wavelength  & DIT\tablefootmark{a} & N\tablefootmark{b} & Seeing  & Baseline & PA  \\
Observation  & Start & End                  & array          & mode\tablefootmark{c} & range       &        &   &       &          &      \\
date         &       &                      &                &          & ($\mu$m)    & (s)    &   &($\arcsec$) &    (m)   & ($\degr$)  \\
\noalign{\smallskip}
\hline
\noalign{\smallskip}
2013 Feb. 28 & 08:30 & 08:55                & UT1-UT2-UT3    &  MR-K-2.1  & 1.926--2.275 & 0.3       & 3200     & 0.8-1.1 & 40/46/85 & 54/38/45  \\
2013 Feb. 28 & 09:11 & 09:42                & UT1-UT2-UT3    &  MR-K-2.1  & 1.926--2.275 & 0.3       & 4800     & 0.6-0.9 & 38/44/81 & 62/45/53  \\
\noalign{\smallskip}
\hline
\end{tabular}
%
\end{scriptsize}
\begin{flushleft}
\hspace{0mm}
\tablefoot{
\tablefoottext{a}{Detector integration time per interferogram.}
\tablefoottext{b}{Number of interferograms.}
}
\end{flushleft}
\end{table*}


\begin{figure*}[h!]

        \includegraphics[width= 18.8 cm]{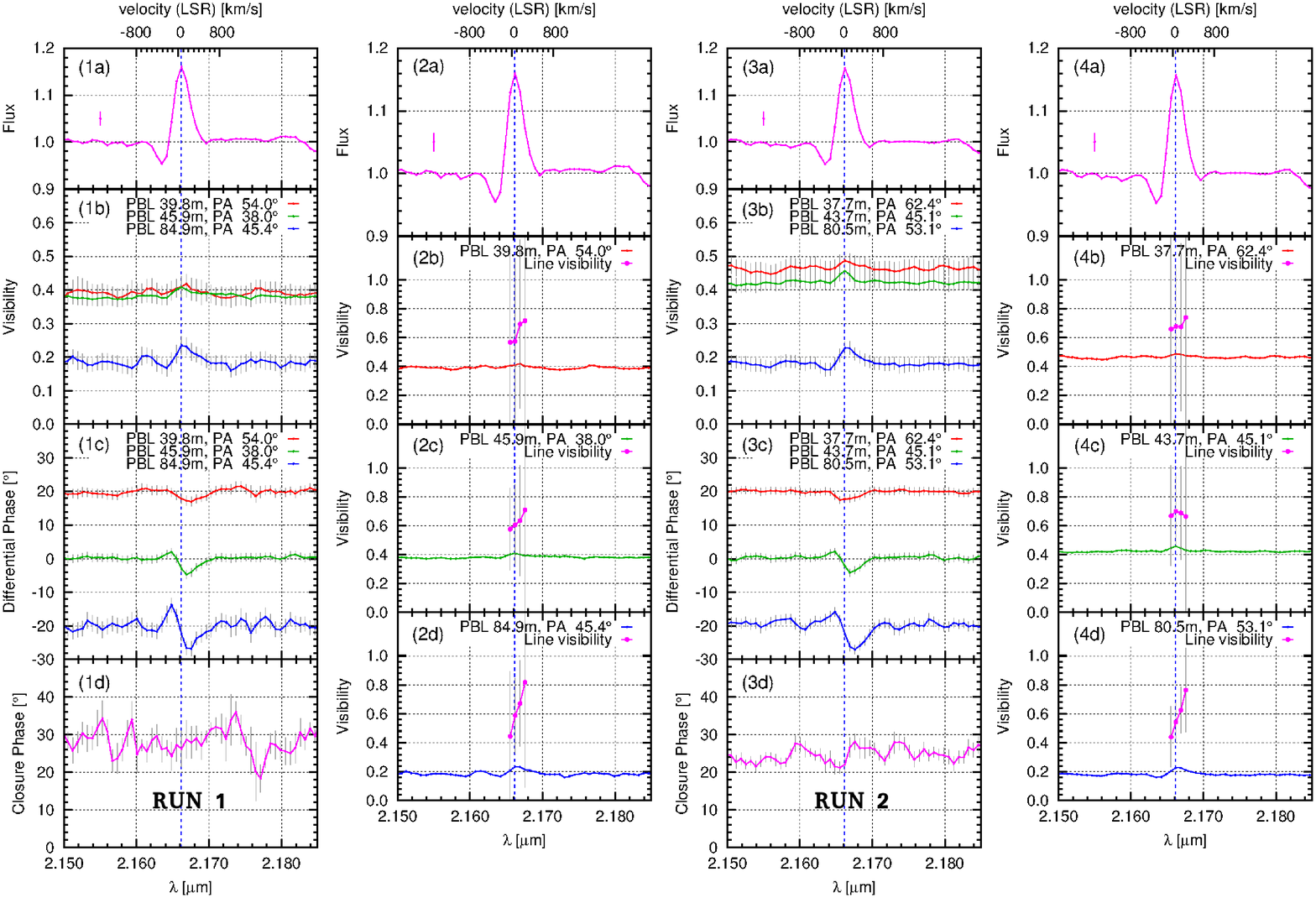}

                \caption{\label{fig:interferometry} {\it Left: Panel\,1.} AMBER-MR interferometric measurements of the Br$\gamma$ line in IRAS\,13481-6124 for run 1 (inserts 1a--1d).
                From  top to bottom: line flux (1a), wavelength-dependent visibilities (1b), differential phases (1c), and closure phase (1d), 
                observed at three different projected baselines (see labels in figure).
                For clarity, the differential phases of the first and last baselines are shifted by +20$\degr$ and -20$\degr$, respectively.
                {\it Middle left: Panel\,2.} Visibilities and continuum-corrected (pure) Br$\gamma$-line visibilities of our AMBER-MR observation of IRAS\,13481-6124 for run 1 (inserts 2b--2d).
                From top to bottom: line flux (2a), visibilities of first (2b), second (2c) and third baseline (2d). 
                {\it Middle right: Panel\,3.} AMBER-MR interferometric measurements of the Br$\gamma$ line in IRAS\,13481-6124 for run 2 (inserts 3a--3d).
                {\it Right: Panel\,4.} Visibilities and continuum-corrected (pure) Br$\gamma$-line visibilities of our AMBER-MR observation of IRAS\,13481-6124 for run 2 (inserts 4b--4d).
                Blue dashed lines encompass the emission peak of the Br$\gamma$ line.}

\end{figure*}



\section{Results}
\label{results:sec}

Our AMBER-MR spectrum shows a rising continuum and a bright Br$\gamma$ emission line with a P\,Cygni profile (inserts 1a--4a of Figure~\ref{fig:interferometry}).
No other lines above a three sigma threshold are detected in the spectrum.  
Figure~\ref{fig:interferometry} shows our interferometric observables (line profile - inserts 1a--4a; visibilities - inserts 1b and 3b; differential phases - inserts 1c and 3c, closure phase
- inserts 1d and 3d) of the Br$\gamma$ line and adjacent continuum 
for the first (Panel\,1) and second run (Panel\,3) along with the inferred continuum-corrected Br$\gamma$-line visibilities (inserts 2b--2d and 4b--4d) 
in four spectral channels (namely those with line-to-continuum ratio larger or equal to 1.1) of run 1 (Panel\,2) and run 2 (Panel\,4). 
Visibilities give information on the size of the emitting region (continuum and/or line), whereas the closure phase quantifies its asymmetry. 
Differential phases are a measure of the photocentre shift of the line with respect to the continuum.

The upper inserts of Figure~\ref{fig:interferometry} (1a--4a) display the Br$\gamma$-line profile normalised to the continuum.
The line is spectrally resolved and shows a wide range of velocities (from $\sim$-500 to 500\,km\,s$^{-1}$).
The emission line peaks at about 50\,km\,s$^{-1}$, with a broad red-shifted wing extending up to $\sim$500\,km\,s$^{-1}$.
The absorption feature on the blue-shifted side ranges from about -500 to -150\,km\,s$^{-1}$. 
The P\,Cygni profile indicates the presence of outflowing matter in the form of a stellar wind or jet, which produces
the typical blue-shifted absorption feature through self-absorption~\citep[see, e.g.][]{mitchell}. The outflow terminal velocity is $\sim$-500\,km\,s$^{-1}$.

In Fig.~\ref{fig:interferometry}, upper-middle (1b and 3b) and lower-middle (1c and 3c) inserts of Panel\,1 (left) and Panel\,3 (middle-right)  
show the wavelength-dependent visibilities and differential phases for the three observed baselines of run 1 (Panel\,1) and run 2 (Panel\,1), respectively,
whereas the closure phases are shown in the lower inserts (1b and 3b). The results of the two runs are similar although,
as mentioned in Sect.~\ref{observations:sec}, the second run data have higher S/N ratio. 
Across the line profile, changes in the visibility and differential phase with respect to the continuum are observed.
Blue dashed lines depict the position of the Br$\gamma$-emission peak in the four panels.

The Br$\gamma$ visibility at the line peak is larger than the continuum visibility at all  six baselines
(see middle-upper panels of left and middle-right inserts in Fig.~\ref{fig:interferometry}), 
indicating that, on average, the Br$\gamma$-emitting region is spatially resolved and more compact than the continuum.
The absolute visibilities of the shortest baselines (red and green lines) display higher values, which indicates that both continuum and line are less spatially resolved along
these baselines with respect to the longest ones. These short baselines are comparable in length and thus the wavelength-dispersed visibilities are very similar. Notably, 
the values of the visibility across the Br$\gamma$-absorption feature at the two longest baselines (85 and 81\,m, P.A. = 45$\degr$ and 53$\degr$, respectively) 
are slightly smaller than the average continuum visibility. 
This means that the Br$\gamma$ photons at high blue-shifted velocities must originate from a region larger than the continuum, whereas the red-shifted Br$\gamma$ emission must 
come from a region that is smaller than the continuum. In other words, the size of the blue-shifted gas in absorption is larger than that of the emitting gas,
suggesting the presence of a wind or an outflow.

The differential phase (DP) at the four longest baselines displays an `S' shape, more pronounced at the 85 and 81\,m baselines, whereas 
the differential phase at the shortest baselines (40 and 38\,m, P.A. = 62$\degr$ and 54$\degr$, respectively) shows a significant displacement with respect to the continuum that appears only 
close to the line peak and along the red-shifted wing (see middle-lower panels of left and middle-right inserts in Fig.~\ref{fig:interferometry}).
This indicates that the outflow is detected mainly at the longest baselines and at P.A.s up to 53$\degr$, i.e. that the outflow is quite collimated (within $\sim$27$\degr$ from the
jet axis, P.A.~206$\degr$/26$\degr$). 
Moreover, since the differential phases are related to the line photocentre shifts with respect to the continuum,
our findings indicate that both red-shifted and blue-shifted line emission wings, at the longest baselines, are spatially extended and located in opposite directions with respect to the continuum emission.

We note that the observed closure phase (CP) (see lower panels of left and middle-right inserts in Fig.~\ref{fig:interferometry}) differs from zero. 
Inside the error bars,
no significant closure-phase variations of the Br$\gamma$ line, with respect to the continuum, are detected.
This is because of the high noise level in the wavelength differential CPs ($\sim$10$\degr$),
which is about one order of magnitude higher than that in the DPs.
The non-zero closure phase originates from the asymmetry of the brightness distribution of the continuum.
Owing to the disc inclination of $\sim$45$\degr$ with respect to the plane of the sky, the far side of the inner disc rim (i.e. the one towards the blue-shifted side of the jet)
displays an area larger (i.e. brighter) than
the near side~\citep[see panel\,d of Fig.~1 in][]{kraus10}. To confirm this scenario, we compare our observations 
with the synthetic images around 2.16\,$\mu$m that are derived from the radiative transfer model presented by \citet{kraus10}.
Although this model was originally adjusted to fit AMBER visibilities and closure phases taken with the auxiliary telescopes, 
it also provides a very good fit to our new, higher SNR UT data. The predicted CP value of $\sim$26$\degr$ is comparable to the measured
values of 25$\degr$$\pm$4$\degr$ (run 2) and 29$\degr$$\pm$6$\degr$ (run 1).
Notably the continuum asymmetry also affects the observed DPs of the Br$\gamma$ line, causing the blue-shifted DPs to be systematically smaller than the red-shifted ones.

\begin{figure}
\centering
\includegraphics[width=9.0cm]{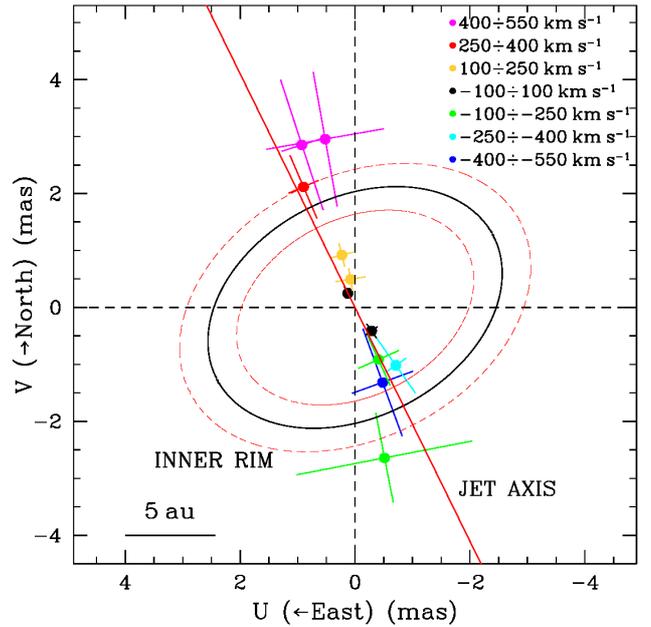}
\caption{\label{fig:UV} Br$\gamma$ displacement for different (radial) velocity channels (with velocity bins from -550 to 550\,km\,s$^{-1}$) derived from the 
measured differential phases, after correcting for the continuum contribution. Different colours indicate different velocity bins. 
Black solid and red dashed ellipses indicate the extent of the K-band continuum with its uncertainty~\citep[1.94$\pm$0.37\,mas][]{kraus10},
whereas the red line shows the position of the jet axis~\citep[][]{caratti15}.}
\end{figure}
  

To infer the size of the Br$\gamma$-emitting region across the different velocity channels, we first compute the continuum-subtracted (or pure-line) 
visibilities at the six baselines (see middle-left and right inserts in Fig.~\ref{fig:interferometry}), following \citet{weigelt07}. 
The ring-fit diameter of the continuum is taken from \citet{kraus10} (5.4\,mas or 17.3\,au at 3.2\,kpc).
The average diameter of the Br$\gamma$-emitting region (averaged over 8 channels, 4 for each baseline, as plotted for the pure line visibilities of Fig.~\ref{fig:interferometry}) 
is $\sim$4\,mas (i.\,e. $\sim$13\,au at 3.2\,kpc) at the two shortest baselines
(37.7 and 39.8\,m), $\sim$3.4\,mas ($\sim$11\,au) at the medium baselines (43.7 and 45.9\,m), 
and $\sim$2\,mas (6.4\,au) at the longest baselines (80.5 and 84.9\,m). On average the size of the Br$\gamma$-emitting region is smaller
than the continuum-emitting region. 
It is worth noting, however, that 
the P\,Cygni absorption in the blue wing of the Br$\gamma$ line makes
the continuum correction of the visibility uncertain for wavelengths
shorter than the peak wavelength of the emission
line. In this case, the continuum-correction  overestimates
the size of the line-emitting region. This might explain why the values of the pure line visibilities (middle-left and right inserts of Fig.~\ref{fig:interferometry})
decrease, moving from red-shifted to blue-shifted wavelengths. 
On the other hand, if this effect is real, it would imply that the two lobes are not spatially symmetric, probably because of the screening effect of the disc 
on the red-shifted lobe of the outflow. 

Finally, the calibrated DPs were converted into photocentre shifts (p) of the Br$\gamma$-velocity components, with respect to the continuum emission,
by solving the following equation: $p_i = (-DP_i \lambda)/(2 \pi B_i)$, where $p_i$ is
the photocentre displacement projection of the 2D photocentre vector $\vec{p}$ on the baseline $B_i$, and $\lambda$ is 
the wavelength of the considered spectral channel~\citep[][]{lebouquin}.
A single astrometric solution (i.e. a single 2D vector $\vec{p}$) was fitted to all baselines within each spectral channel. 
As for the visibilities, we then subtracted the continuum contribution to get
the photocentre displacements of the pure-line emitting region~\citep[see Eq.~3 in][]{kraus12a}. Figure~\ref{fig:UV} shows the continuum-subtracted Br$\gamma$-line photocentre displacements (with the
errorbars in polar coordinates), which were 
derived from the astrometric solution in the different (radial) velocity channels
(colour coded). Both the inner rim, defined by the study of the NIR continuum~\citep[see][]{kraus10}, and the known jet axis~\citep[see][]{caratti15}, are shown as a black solid ellipse and red solid line, respectively.
The red dashed ellipses indicate the uncertainty on the size of the inner rim~\citep[1.97$\pm$0.37\,mas;][]{kraus10}.
Fig.~\ref{fig:UV} shows that the photocentre-shifts in the blue-shifted (green, cyan and blue dots; from ${\rm v}_{\rm r}\sim$-420 to $\sim$-140\,km\,s$^{-1}$) and red-shifted 
(orange, red, and magenta dots; from ${\rm v}_{\rm r} \sim$140 to $\sim$520\,km\,s$^{-1}$) wings of the Br$\gamma$ are  increasing considerably with increasing velocities and are roughly displaced along a straight line, 
close to the jet axis, in the blue- and red-shifted lobes, respectively. 
The significance of the red-shifted data points at the highest velocity (magenta dots) is reduced with respect to the other points since they are seen out of the obscuring 
disc only because of the large uncertainty of their position.

The low-velocity components (black dots; with ${\rm v}_{\rm r} \sim$-50 and 50\,km\,s$^{-1}$) are also aligned with the jet axis, but are much closer to the central source ($\lesssim$2\,au),
indicating a strong acceleration of the gas in the inner 10\,au. We note that,
at 2\,au, the Keplerian velocity of a 20\,M$_\sun$ object is $\sim$95\,km\,s$^{-1}$ (or ${\rm v}_{\rm r}\sim$67\,\,km\,s$^{-1}$), that is, 
it is not spectrally resolved by our observations ($\Delta$v $\sim$ 140\,km\,s$^{-1}$). 
Our findings thus suggest that, on the one hand, the high-velocity component is tracing the jet/outflow, 
and, on the other, the low-velocity component is tracing slow moving gas, such as a disc wind (in Keplerian rotation) or
the very base of the jet, namely the region of the disc where the jet footpoints are located.

\section{Origin of the Br$\gamma$ line in IRAS\,13481-6124}
\label{discussion:sec}

The origin of the Br$\gamma$ line in YSOs is controversial and there are
several physical mechanisms that could produce this emission, such as accretion of matter onto the star~\citep[e.g.][]{eisner09},
or outflowing material from a disc wind~\citep[e.g.][]{weigelt,rebeca15,caratti15b}, extended wind or outflow~\citep[e.g.][]{rebeca16}, or jet~\citep[e.g.][]{stecklum12}.

Our AMBER/VLTI observations of IRAS\,13481-6124 suggest that the main Br$\gamma$ emission emanates from an ionised jet.
The observed Br$\gamma$ P\,Cygni profile indicates the presence of a fast ionised jet/outflow ($\sim$500\,km\,s$^{-1}$).
This value agrees well with the radio jet velocities measured in other HMYSOs~\citep[see, e.g.][]{heathcote98,marti,curiel,torrelles11}.
Visibilities and differential phases at high velocities indicate that the outflowing matter is spatially extended, from a few au to a few tens of au, and the photocentre shifts 
grow with increasing radial velocities.
Both observables also suggest that the flow is relatively well collimated ($\lesssim$30$\degr$) close to the known parsec-scale jet P.A.,
implying that we are observing a collimated wind or flow.
Initial jet opening angles are typically $\sim$30$\degr$ in YSOs and a considerable degree of focusing in low-mass YSOs happens at several au from the source~\citep[see, e.g.][]{ray07}.
It is thus reasonable to presume that the corresponding jet collimation in HMYSOs occurs at even larger distances from the source (a few tens of au).
Indeed, as magneto-centrifugal launching models are scale-free~\citep[see, e.g.][]{ferreira97,ferreira04,pudritz_PPV}, for higher masses 
the configuration readjusts to larger scales consistently with the different values of the gravitational potential.
As IRAS\,13481-6124 has a parsec-scale jet, we could then argue that the high-velocity component of the Br$\gamma$ line (500 < ${\rm v}_{\rm r}$ < 100\,km\,s$^{-1}$) 
is tracing the (poorly collimated) jet, which extends from a few au to tens of au from the central source.
On the other hand, the low-velocity component (${\rm v}_{\rm r}$ < 100\,km\,s$^{-1}$),
although partially resolved, is more compact, being located $\lesssim$2\,au from the source and 
well inside the inner-rim disc. This slow moving gas component is also aligned with the jet axis.
Owing to our limited spectral resolution and given that this velocity is also compatible with Keplerian rotation 
(at a distance of $\lesssim$2\,au), this component might originate from a disc wind or the jet foot-point. 
We note that a similar geometry for the Br$\gamma$ emission in another HMYSO (\object{W 33A}) was inferred by \citet{davies} with spectro-astrometry.
Finally, even if the error bars are considered, the Br$\gamma$ visibility is lower than one, i.e. it is spatially resolved at the longest baselines, indicating that the
bulk of the emission cannot originate from accretion, which would not be spatially resolved at our baselines.
Moreover, infall gas would have free-fall velocities of several hundreds of km\,s$^{-1}$, which are not detected close to the source.

In conclusion, most of the observed Br$\gamma$ emission must originate from the ionised jet.
In principle, it should be fully ionised because  of being
exposed to the UV radiation of a 20\,M$_\sun$ HMYSO~\citep[][]{tanaka}. 
As for the case of irradiated jets in massive star-forming regions~\citep[see, e.g.][]{reipurth98,bally06},
the gas of the jet would then be fully traced by the HI emission and its denisty proportional to the HI line intensity.
Therefore, measurements of intensity, velocity, and size of the Br$\gamma$-emitting region might provide a good estimate
of the mass-loss rate.

\begin{acknowledgements}
A.C.G., R.G.L., and T.P.R. were supported by Science Foundation Ireland, grant 13/ERC/I2907.
A.K. and S.K. acknowledge support from a STFC Ernest Rutherford fellowship and grant (ST/J004030/1, ST/K003445/1), and Marie-Sklodowska Curie CIG grant (Ref. 618910).
A.S. was supported by the Deutsche
Forschungsgemeinschaft (DFG) Priority Program 1573.
This research has also made use of NASA's Astrophysics Data System Bibliographic Services and the SIMBAD database operated
at the CDS, Strasbourg, France.
\end{acknowledgements}

\bibliographystyle{aa}
\bibliography{references}
\end{document}